\documentclass[aps,twocolumn,preprintnumbers,amsmath,amssymb,superscriptaddress,floatfix,nofootinbib]{revtex4}
\usepackage{graphicx,color,dcolumn,booktabs,bm}
\usepackage{longtable,lscape}
\usepackage{txfonts}
\usepackage{overpic}
\usepackage{epsfig}
\usepackage{amssymb}
\usepackage{rotating}
\usepackage{epstopdf}
\usepackage{indentfirst}
\usepackage{feynmf}   %{feynmp}
\usepackage{slashed}  %for Feynman symbols
\usepackage{cases}
\usepackage{color}
\usepackage{float}
\usepackage{multirow}
\usepackage{graphicx,color,dcolumn,booktabs,bm}
\usepackage{cases}
\usepackage{array}

\graphicspath{{Figures/}} %

\usepackage[colorlinks, citecolor=blue,anchorcolor=red,menucolor=red,
linkcolor=red,filecolor=red,runcolor=red,urlcolor=blue,frenchlinks=red]{hyperref}

\begin{document}

\title{Masses of fully heavy tetraquarks $QQ \bar Q \bar Q$ in an extended relativized quark model}

\author{Qi-Fang L\"u} \email{lvqifang@hunnu.edu.cn}
\affiliation{  Department
of Physics, Hunan Normal University,  Changsha 410081, China }

\affiliation{ Synergetic Innovation
Center for Quantum Effects and Applications (SICQEA), Changsha 410081,China}

\affiliation{  Key Laboratory of
Low-Dimensional Quantum Structures and Quantum Control of Ministry
of Education, Changsha 410081, China}
\author{Dian-Yong Chen} \email{chendy@seu.edu.cn}
\affiliation{School of Physics, Southeast University, Nanjing 210094, China }
\author{Yu-Bing Dong} \email{dongyb@ihep.ac.cn}
\affiliation{Institute of High Energy Physics, Chinese Academy of Sciences, Beijing 100049, China}
\affiliation{Theoretical Physics Center for Science Facilities (TPCSF), CAS, Beijing 100049, China}
\affiliation{School of Physical Sciences, University of Chinese Academy of Sciences, Beijing 101408, China}

\begin{abstract}
Inspired by recent measurement of possible fully charmed tetraquarks in LHCb Collaboration, we investigate the mass spectra of fully heavy tetraquarks $QQ \bar Q \bar Q$ in an extended relativized quark model. Our estimations indicate that the broad structure around 6.4 GeV should contain one or more ground $cc \bar c \bar c$ tetraquark states, while the narrow structure near 6.9 GeV can be categorized  as the first radial excitation of $cc \bar c \bar c$ system. Moreover, with the wave functions of the tetraquarks and mesons, the strong decays of tetraquarks into heavy quarkonium pair are qualitatively discussed, which can be further checked by the LHCb and CMS Collaborations.
\end{abstract}

\maketitle

\section{Introduction}{\label{Sec:Int}}

Since the observation of $X(3872)$ in 2003~\cite{Choi:2003ue}, the searching for hadrons beyond the conventional mesons and baryons have become one of  intriguing topics in the past decades. On the experimental side, a growing number of new hadron states have been observed experimentally. Some of these states cannot be accommodated into the traditional mesons or baryons, which can be good candidates of molecular or tetraquark states. Recent experimental and theoretical status can be found in the literature reviews~\cite{Klempt:2007cp, Brambilla:2010cs, Chen:2016qju, Lebed:2016hpi,  Guo:2017jvc, Esposito:2016noz,Ali:2017jda, Liu:2019zoy,Brambilla:2019esw,Dong:2017gaw}.

Among the observed new hadron states, those with heavy quark components are particularly interesting, since the spectroscopy of traditional mesons and baryons with heavy quarks are much clear than the light hadrons. Moreover, the interactions involved heavy quarks are supposed to be dominated by the short range one gluon exchange potential rather than the long range potential resulted from light meson exchanges. Thus, the new hadron states composed by four heavy quarks should be good candidates of compact tetraquark states rather than deuteron-like molecular states.     

In 2017, the CMS Collaboration reported their measurement of exotic structure in four lepton channel and found an excess in $18.4\pm 0.1 (\mathrm{stat.}) \pm 0.2 (\mathrm{syst.}) \mathrm{GeV}/c^2$ with a global significance of 3.6 $\sigma$~\cite{Khachatryan:2016ydm}. This structure indicates a possible $bb\bar{b}\bar{b}$ tetraquark state~\cite{CMSMeeting, Khachatryan:2016ydm, Yi:2018fxo}. It should be noticed that this structure is below the threshold of bottomonium meson pair, which demonstrates that the decays into bottomonium meson pair through quark rearrangement should be hindered. Later, the LHCb and CMS Collaborations analyzed the invariant mass distributions of $\Upsilon(1S) \mu^+ \mu^-$, but no evident structure was observed~\cite{Aaij:2018zrb,Sirunyan:2020txn}.

On the theoretical side, the compact tetraquark states composed of $bb\bar{b}\bar{b}$ have been investigated extensively, but the conclusions are model dependent. In Refs.~\cite{Wang:2017jtz,Karliner:2016zzc,Berezhnoy:2011xn,Bai:2016int,Anwar:2017toa,Esposito:2018cwh,Chen:2016jxd,Debastiani:2017msn,Wang:2018poa}, the lowest $bb\bar{b}\bar{b}$ tetraquark state is estimated to be below the threshold of bottomonium meson pair, while in Refs.~\cite{Wu:2016vtq,Lloyd:2003yc,Ader:1981db,Hughes:2017xie,Richard:2018yrm, Liu:2019zuc, Wang:2019rdo,Chen:2019dvd,Deng:2020iqw}, all the $bb\bar{b}\bar{b}$ tetraquark states are above the threshold. To further distinguish different model and reveal the underlying dynamics of fully heavy tetraquark states, more efforts are needed, especially from the experimental side.

Very recently, the LHCb Collaboration reported their measurement of the $J/\psi$ pair invariant mass spectrum and a structure near 6.9 GeV/$c^2$ was observed with the significance greater than $5\sigma$~\cite{LHCb:2020}. The resonance parameters are fitted to be 
\begin{eqnarray}
	m &=& 6905 \pm 11(\mathrm{stat.}) \pm 7 (\mathrm{syst.})  \ \mathrm{MeV}/c^2 \nonumber \\
    \Gamma	&=& 80 \pm 19(\mathrm{stat.}) \pm 33(\mathrm{syst.})  \ \mathrm{MeV}/c^2,
\end{eqnarray}
in a no-interference scenario, or 
\begin{eqnarray}
	m &=&6886 \pm 11 (\mathrm{stat.}) \pm 11(\mathrm{syst.})  \ \mathrm{MeV}/c^2 \nonumber \\
    \Gamma	&=& 168 \pm 33(\mathrm{stat.}) \pm \mathrm{69}(\mathrm{syst.})\ \mathrm{MeV}/c^2,
\end{eqnarray}
in an interference scenario. Besides the structure near 6.9 GeV, the experimental data also indicated another two structures in the vicinity of 6.4 GeV and 7.2 GeV, respectively~\cite{LHCb:2020}. These  structures may be the evidence of compact tetraquark state composed by $cc\bar{c}\bar{c}$, which can be a criterion for different models. 

After the observation of the LHCb Collaboration, the state around 6.9 GeV has been investigated in different models. In Ref.~\cite{liu:2020eha}, this state was interpreted as a $P-$wave tetraquark state in a nonrelativistic quark model, while the QCD sum rule estimations indicated that it could be a second radial excited  $S-$wave tetraquark state~\cite{Wang:2020ols}. The results in Refs.~\cite{1802716,1802721} suggested that the resonances with  $J^P=0^+$ and $1^+$ are about $6.4 \sim 6.6$ GeV, while the $2^+$ state is about 7.0 GeV, which are consistent with the structures reported by LHCb Collaboration \cite{LHCb:2020}.

In Ref.~\cite{Lu:2020rog}, we extended the relativized quark model proposed by Godfrey and Isgur to investigate the doubly heavy tetraquarks with the same model parameters. With such an extension, the tetraquaks and conventional mesons can be described in a uniform frame. In the present work, we further study the full heavy tetraquarks $QQ\bar{Q}\bar{Q}$ in the extended relativized quark model and give possible interpretation of the newly observed state around 6.9 GeV. Moreover, the newly observed structures are above the threshold of heavy quarkonium pair, thus, these states can decay into heavy quarkonium pair by quark rearrangement. For simplicity, the decay amplitude should be proportional to the overlap of wave functions of the initial and final states, thus, we can qualitatively discuss the decay behaviors of tetraquarks with the wave functions estimated from the relativized quark model. 

This work is organized as follows. In section~\ref{model}, we present a review of the extended relativized quark model used in the present work. The numerical results of the masses and decays for the tetraquarks are given in Section~\ref{results}. The last section is devoted to a brief summary.

\section{Extended relativized quark model}{\label{model}}

To investigate the masses of fully heavy tetraquarks $Q_1 Q_2 \bar Q_3 \bar Q_4$, we employ an extended relativized quark model, which has been developed very recently for the tetraquark states~\cite{Lu:2020rog}. It is an extension of the relativized quark model to deal with the four-body systems. The Hamiltonian for a $Q_1 Q_2 \bar Q_3 \bar Q_4$ state can be expressed as
\begin{equation}
H = H_0+\sum_{i<j}V_{ij}^{\rm oge}+\sum_{i<j}V_{ij}^{\rm conf}, \label{ham}
\end{equation}
where 
\begin{equation}
H_0 = \sum_{i=1}^{4}(p_i^2+m_i^2)^{1/2}
\end{equation}
is the relativistic kinetic energy, $V_{ij}^{\rm oge}$ is the one gluon exchange potential including the spin-spin interaction, and $V_{ij}^{\rm conf}$ stands
for the confining part. The explicit formula and parameters of relativized potentials can be found in Refs.~\cite{Godfrey:1985xj,Lu:2020rog}.

The wave function of a $Q_1 Q_2 \bar Q_3 \bar Q_4$ state is composed of color, flavor, spin, and spatial parts.
In the color space, two types of colorless states with determinate permutation properties exist
\begin{equation}
|\bar 3 3\rangle = |(Q_1 Q_2)^{\bar 3} (\bar Q_3 \bar Q_4)^3\rangle,
\end{equation}
\begin{equation}
|6 \bar 6\rangle = |(Q_1 Q_2)^{6} (\bar Q_3 \bar Q_4)^{\bar 6}\rangle,
\end{equation}
where the $|\bar 3 3\rangle$ and $|6 \bar 6 \rangle$ are antisymmetric and symmetric under the exchange of $Q_1Q_2$ or $\bar Q_3 \bar Q_4$, respectively. In the flavor space, the combinations of $\{ c c \}$, $\{ \bar c \bar c \}$, $\{b b \}$, and $\{ \bar b \bar b \}$ are
always symmetric, where the braces $\{ ~ \}$ are adopted to stand for symmetric flavor wave functions.

For the spin part, the six spin bases can be written as,
\begin{equation}
\chi^{00}_0 = |(Q_1 Q_2)_0 (\bar Q_3 \bar Q_4)_0\rangle_0,
\end{equation}
\begin{equation}
\chi^{11}_0 = |(Q_1 Q_2)_1 (\bar Q_3 \bar Q_4)_1\rangle_0,
\end{equation}
\begin{equation}
\chi^{01}_1 = |(Q_1 Q_2)_0 (\bar Q_3 \bar Q_4)_1\rangle_1,
\end{equation}
\begin{equation}
\chi^{10}_1 = |(Q_1 Q_2)_1 (\bar Q_3 \bar Q_4)_0\rangle_1,
\end{equation}
\begin{equation}
\chi^{11}_1 = |(Q_1 Q_2)_1 (\bar Q_3 \bar Q_4)_1\rangle_1,
\end{equation}
\begin{equation}
\chi^{11}_2 = |(Q_1 Q_2)_1 (\bar Q_3 \bar Q_4)_1\rangle_2,
\end{equation}
where $(Q_1 Q_2)_0$ and $(\bar Q_3 \bar Q_4)_0$ are antisymmetric and the
$(Q_1 Q_2)_1$ and $(\bar Q_3 \bar Q_4)_1$ are symmetric for the two fermions under permutations. The matrix elements of the color and spin parts are same as the
doubly heavy tetraquarks~\cite{Lu:2020rog}.

\begin{figure*}[!htbp]
\includegraphics[scale=0.7]{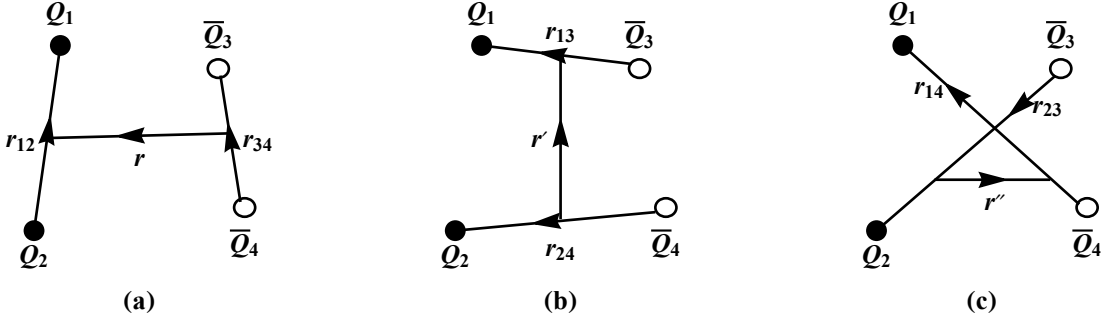}
\vspace{0.0cm} \caption{The $Q_1Q_2 \bar Q_3 \bar Q_4$ tetraquark state in Jacobi coordinates.}
\label{jacobi}
\end{figure*}

In the spatial space, the Jacobi coordinates are presented in Figure~\ref{jacobi}. For the $cc \bar c \bar c$ and $bb \bar b \bar b$ systems,
we can define
\begin{equation}
\boldsymbol r_{12}=  \boldsymbol r_1 - \boldsymbol r_2,
\end{equation}
\begin{equation}
\boldsymbol r_{34}=  \boldsymbol r_3 - \boldsymbol r_4,
\end{equation}
\begin{equation}
\boldsymbol r = \frac{\boldsymbol r_1 + \boldsymbol r_2}{2} - \frac{\boldsymbol r_3 + \boldsymbol r_4}{2},
\end{equation}
and
\begin{equation}
\boldsymbol R = \frac{\boldsymbol r_1 + \boldsymbol r_2
+ \boldsymbol r_3 + \boldsymbol r_4}{4}.
\end{equation}
Then, other relevant coordinates of this system can be obtained in terms of $\boldsymbol r_{12}$, $\boldsymbol r_{34}$,
and $\boldsymbol r$. For a $S-$wave state, we adopt a set of Gaussian
functions to approach its realistic spatial wave function~\cite{Hiyama:2003cu}
\begin{equation}
\Psi(\boldsymbol r_{12},\boldsymbol r_{34},\boldsymbol r) = \sum_{n_{12},n_{34},n} C_{n_{12}n_{34}n}
\psi_{n_{12}}(\boldsymbol r_{12}) \psi_{n_{34}}(\boldsymbol r_{34}) \psi_n(\boldsymbol r),
\end{equation}
where $C_{n_{12}n_{34}n}$ are the expansion coefficients. The $\psi_{n_{12}}(\boldsymbol r_{12}) \psi_{n_{34}}(\boldsymbol r_{34})
\psi_n(\boldsymbol r)$ is the position representation of the basis $|n_{12}n_{34}n\rangle$, where
\begin{equation}
\psi_n(\boldsymbol r) = \frac{2^{7/4}\nu_n^{3/4}}{\pi^{1/4}} e^{-\nu_n r^2} Y_{00}(\hat{\boldsymbol r}) =
\Bigg(\frac{2 \nu_n}{\pi} \Bigg )^{3/4} e^{-\nu_n r^2},
\end{equation}
\begin{equation}
\nu_n = \frac{1}{r_1^2a^{2(n-1)}},~~~~ (n=1-N_{max}).
\end{equation}
It should be stressed that our final results are independent on geometric Gaussian size parameters $r_1$, $a$, and $N_{max}$ when 
sufficiently large bases are chosen~\cite{Hiyama:2003cu}. The $\psi_{n_{12}}(\boldsymbol r_{12})$ and  $\psi_{n_{34}}(\boldsymbol r_{34})$
can be written in a similar way, and the momentum representation the basis $ |n_{12}n_{34}n\rangle$ can be obtained
by the Fourier transformation.

According to the Pauli exclusion principle, the total wave function of a tetraquark should be antisymmetric, and possible configurations for
$cc \bar c \bar c$ and $bb \bar b \bar b$ systems are presented in Table~\ref{configuration}. With the full wave functions, all the matrix elements of the Hamiltonian can be worked out. Then, the masses can be obtained by solving the following generalized eigenvalue problem
\begin{equation}
\sum_{j=1}^{N_{max}^3}(H_{ij}-EN_{ij})C_j=0,~~~~ (i=1-N_{max}^3),
\end{equation}
where the $H_{ij}$ are the matrix elements in the total bases, $N_{ij}$ is the overlap matrix elements of the Guassian functions arising form
the nonorthogonality of bases, $E$ stands for the mass, and $C_j$
are the eigenvector corresponding to the coefficients $C_{n_{12}n_{34}n}$ of spatial wave function. Moreover, for a given system, different configurations with same $J^{PC}$ can mix with each other. The mixing effects are taken into account by diagonalizing the mass matrix of these configurations.

\begin{table}[!htbp]
\begin{center}
\caption{ \label{configuration} Possible configurations for the $cc \bar c \bar c$ and $bb \bar b \bar b$ systems.
The subscripts and superscripts denote the
spin and color quantum numbers, respectively.}
\begin{tabular*}{8.6cm}{@{\extracolsep{\fill}}*{4}{p{2cm}<{\centering}}}
\hline\hline
System & $J^P$         &  \multicolumn{2}{c}{Configuration} \\\hline
$ cc \bar c \bar c$ & $0^{++}$       &  $|\{cc\}^{\bar 3}_1 \{\bar c \bar c\}^3_1\rangle_0$
&  $|\{cc\}^6_0 \{\bar c \bar c\}^{\bar 6}_0\rangle_0$      \\
& $1^{+-}$       &  $|\{cc\}^{\bar 3}_1 \{\bar c \bar c\}^3_1\rangle_1$     &  $\cdots$    \\
& $2^{++}$       &  $|\{cc\}^{\bar 3}_1 \{\bar c \bar c\}^3_1\rangle_2$     &  $\cdots$     \\
$bb \bar b \bar b $ & $0^{++}$       &  $|\{bb\}^{\bar 3}_1 \{\bar b \bar b\}^3_1\rangle_0$
 &  $|\{bb\}^6_0 \{\bar b \bar b\}^{\bar 6}_0\rangle_0$      \\
& $1^{+-}$       &  $|\{bb\}^{\bar 3}_1 \{\bar b \bar b\}^3_1\rangle_1$      &  $\cdots$     \\
& $2^{++}$       &  $|\{bb\}^{\bar 3}_1 \{\bar b \bar b\}^3_1\rangle_2$     &  $\cdots$   \\\hline
\end{tabular*}
\end{center}
\end{table}

\section{RESULTS AND DISCUSSIONS}{\label{results}}

In present work, we adopt $N^3_{max} = 10^3$ Gaussian bases to estimate the $S-$wave $QQ\bar Q \bar Q$ spectra.  With these large bases,
the numerical results are stable enough for our quark model calculations. The predicted masses of ground states for $cc\bar c \bar c$
and $bb\bar b \bar b$ systems are presented in Table~\ref{mass1}. For the $cc\bar c \bar c$ system, the masses of four ground states lie
in the range $6435\sim 6543 \rm{MeV}$, which are higher than the $J/\psi J/\psi$ threshold.  Compared with the experimental data,
we expect that these states should correspond to the broad structure in the vicinity of 6.4 GeV. This broad structure may be one state
or an overlap of several states from current data, and more experimental information are needed to clarify its nature.
For the $bb\bar b \bar b$ system, the masses are also above the relevant $\Upsilon \Upsilon$ thresholds. These results for the ground states
are consistent with nonrelativistic quark model calculations where the pairwise potentials are adopted
properly~\cite{Liu:2019zuc,Wang:2019rdo,Chen:2019dvd,Richard:2019cmi,Deng:2020iqw}.

\begin{table*}[htbp]
\begin{center}
\caption{\label{mass1} Predicted masses of the ground states for $cc\bar c \bar c$ and $bb\bar b \bar b$ systems.}
\begin{tabular*}{18cm}{@{\extracolsep{\fill}}*{5}{p{3.3cm}<{\centering}}}
\hline\hline
 $J^P$  & Configuration                                             & $\langle H\rangle$ (MeV) & Mass (MeV)  & Eigenvector\\\hline
 $0^{++}$  &  $|\{cc\}^{\bar 3}_1 \{\bar c \bar c\}^3_1\rangle_0$     & \multirow{2}{*}{$\begin{pmatrix}6501&-52 \\-52&6475\end{pmatrix}$}
               & \multirow{2}{*}{$\begin{bmatrix}6435 \\6542 \end{bmatrix}$}  & \multirow{2}{*}{$\begin{bmatrix}(0.617, 0.787)\\
               (0.787, -0.617)\end{bmatrix}$}\\
                 &  $|\{cc\}^6_0 \{\bar c \bar c\}^{\bar 6}_0\rangle_0$    \\
 $1^{+-}$  &  $|\{cc\}^{\bar 3}_1 \{\bar c \bar c\}^3_1\rangle_1$    & 6515  &  6515  &  1\\
 $2^{++}$  &  $|\{cc\}^{\bar 3}_1 \{\bar c \bar c\}^3_1\rangle_2$    & 6543  &  6543  &  1\\\hline

 $0^{++}$  &  $|\{bb\}^{\bar 3}_1 \{\bar b \bar b\}^3_1\rangle_0$     & \multirow{2}{*}{$\begin{pmatrix}19246& 20 \\20&19210\end{pmatrix}$}
               & \multirow{2}{*}{$\begin{bmatrix}19201 \\19255 \end{bmatrix}$}  & \multirow{2}{*}{$\begin{bmatrix}(-0.410, 0.912)\\
               (0.912, 0.410)\end{bmatrix}$}\\
                 &  $|\{bb\}^6_0 \{\bar b \bar b\}^{\bar 6}_0\rangle_0$    \\
 $1^{+-}$  &  $|\{bb\}^{\bar 3}_1 \{\bar b \bar b\}^3_1\rangle_1$    & 19251  &  19251  &  1\\
 $2^{++}$  &  $|\{bb\}^{\bar 3}_1 \{\bar b \bar b\}^3_1\rangle_2$    & 19262  &  19262  &  1\\
\hline\hline
\end{tabular*}
\end{center}
\end{table*}

Besides the masses, we can also calculate the proportions of hidden color components and the root mean square radii.
In addition to $|\bar{3}3 \rangle$ and $|6 \bar{6}\rangle$, other sets of color representations can be defined as
\begin{equation}
|11\rangle = |(Q_1 \bar Q_3)^1 (Q_2 \bar Q_4)^1\rangle,
\end{equation}
\begin{equation}
|88\rangle = |(Q_1 \bar Q_3)^8 (Q_2 \bar Q_4)^8\rangle,
\end{equation}
and
\begin{equation}
|1^\prime 1^\prime\rangle = |(Q_1 \bar Q_4)^1 (Q_2 \bar Q_3)^1\rangle,
\end{equation}
\begin{equation}
|8^\prime 8^\prime\rangle = |(Q_1 \bar Q_4)^8 (Q_2 \bar Q_3)^8\rangle.
\end{equation}
Then, the relations among three sets of color representations can be expressed as follows,
\begin{equation}
|11\rangle = \sqrt{\frac{1}{3}}|\bar 3 3\rangle + \sqrt{\frac{2}{3}}|6 \bar 6\rangle,
\end{equation}
\begin{equation}
|88\rangle = - \sqrt{\frac{2}{3}}|\bar 3 3\rangle + \sqrt{\frac{1}{3}}|6 \bar 6\rangle,
\end{equation}
and
\begin{equation}
|1^\prime 1^\prime\rangle = - \sqrt{\frac{1}{3}}|\bar 3 3\rangle + \sqrt{\frac{2}{3}}|6 \bar 6\rangle,
\end{equation}
\begin{equation}
|8^\prime 8^\prime\rangle = \sqrt{\frac{2}{3}}|\bar 3 3\rangle + \sqrt{\frac{1}{3}}|6 \bar 6\rangle.
\end{equation}
In present work, we adopt the $|11\rangle$ and $|88\rangle$ representations to stand for the neutral color and hidden color
components, respectively. The color proportions and root mean square radii of the calculated ground states are displayed in
Table~\ref{pro1}. For the $cc\bar c \bar c$ and $bb\bar b \bar b$ systems, the expectations satisfy the following relations
\begin{equation}
\langle \boldsymbol r_{12}^2  \rangle^{1/2} = \langle \boldsymbol r_{34}^2  \rangle^{1/2},
\end{equation}
\begin{equation}
\langle \boldsymbol r^{\prime 2}  \rangle^{1/2} = \langle \boldsymbol r^{\prime \prime 2}  \rangle^{1/2},
\end{equation}
\begin{equation}
\langle \boldsymbol r_{13}^2  \rangle^{1/2} = \langle \boldsymbol r_{14}^2  \rangle^{1/2} = \langle \boldsymbol r_{23}^2
\rangle^{1/2} = \langle \boldsymbol r_{23}^2  \rangle^{1/2}.
\end{equation}
From Table~\ref{pro1}, it can be seen that these states have significant hidden color components and small root mean square radii, and this
phenomena indicates that all of them can be regarded as compact tetraquarks.

\begin{table*}[!htb]
\begin{center}
\caption{\label{pro1} The color proportions and the root mean square radii of the ground states for  $cc\bar c \bar c$ and $bb\bar b
\bar b$ systems.  The units of masses and root mean square radii are in MeV and fm, respectively.}
\begin{tabular*}{18cm}{@{\extracolsep{\fill}}*{11}{p{1.3cm}<{\centering}}}
\hline\hline
 System  & $J^P$ & Mass  &   $|\bar 3 3\rangle$  &  $|6 \bar 6\rangle$  &  $|11\rangle$  & $|88\rangle$  &  $\langle \boldsymbol r_{12}^2
 \rangle^{1/2}$     &  $\langle \boldsymbol r^2 \rangle^{1/2}$  &
 $\langle \boldsymbol r_{13}^2  \rangle^{1/2}$
 &  $\langle \boldsymbol r^{\prime 2} \rangle^{1/2}$  \\\hline
 $cc\bar c \bar c$  & $0^{++}$ & 6435  & 38.1\%  &  61.9\% &  54.0\%  & 46.0\%   &  0.433  &  0.265     &  0.405  &  0.306   \\
  & $0^{++}$ & 6542  &   61.9\%  &  38.1\%  &  46.0\%  & 54.0\%   &  0.415  &  0.281     &  0.406  &  0.293  \\
  & $1^{+-}$ & 6515  &   100\%  &  0\%  &  33.3\%  & 66.7\%   &  0.387  &  0.310     &  0.414  &  0.274  \\
  & $2^{++}$ & 6543  &   100\%  &  0\%  &  33.3\%  & 66.7\%   &  0.394  &  0.321     &  0.425  &  0.278  \\\hline
 $bb\bar b \bar b$  & $0^{++}$ & 19201  &   16.8\%  &  83.2\%  &  61.1\%  & 38.9\%   &  0.286  &  0.163     &  0.260  &  0.202  \\
  & $0^{++}$ & 19255  &   83.2\%  &  16.8\%  &  38.9\%  & 61.1\%   &  0.257  &  0.194     &  0.266  &  0.182  \\
  & $1^{+-}$ & 19251  &   100\%  &  0\%  &  33.3\%  & 66.7\%   &  0.251  &  0.203     &  0.269  &  0.177   \\
  & $2^{++}$ & 19262  &   100\%  &  0\%  &  33.3\%  & 66.7\%   &  0.253  &  0.206     &  0.272  &  0.179  \\
\hline\hline
\end{tabular*}
\end{center}
\end{table*}

\begin{figure*}[!htb]
\includegraphics[scale=0.55]{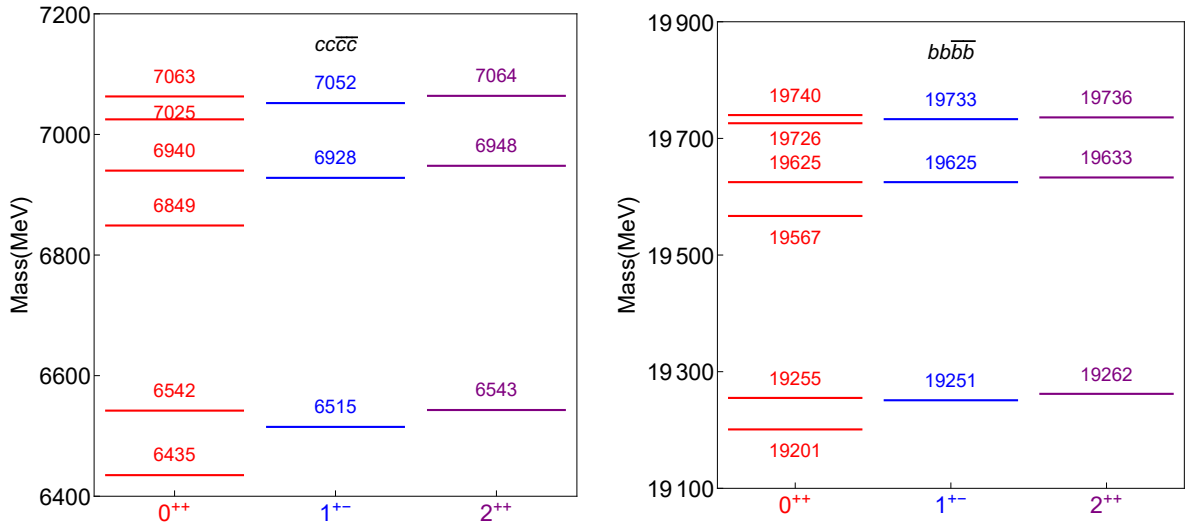}
\vspace{-0.0cm} \caption{The predicted masses of $cc\bar c \bar c$ and  $bb\bar b \bar b$ systems.}
\label{mass}
\end{figure*}

The low-lying radial excitations of $cc\bar c \bar c$ and $bb\bar b \bar b$ systems are also calculated in our approach and the
results are presented in Table~\ref{mass2}. Theoretically, there are two types of radial excitations, the $QQ$ or $\bar Q \bar Q$ mode,
and the one between $QQ$ and $\bar Q \bar Q$ subsystems. The physical states should correspond to the mixture of these two modes.
Our results show that the first excitations for the $cc\bar c \bar c$ states lie around 6900 MeV, which should correspond to the
observed structure near 6.9 GeV by LHCb Collaboration.  Given the $J/\psi J/\psi$ decay mode, the $J^{PC}$ of this structure should
equal to $0^{++}$ or $2^{++}$. Current information is insufficient to determine its spin-parity, and more theoretical and experimental
efforts are needed. Moreover, we find that another set of excitations are around 7050 MeV, which have not been observed by LHCb
Collaboration. In addition, the predicted excitations for the $bb \bar b \bar b$ systems are around 19600 and 19730 MeV, respectively,
which can easily fall apart into the bottomonium pairs. It is claimed that there is no signal in the $\Upsilon \mu^+ \mu^-$
channels by previous LHCb and CMS experiments~\cite{Aaij:2018zrb,Sirunyan:2020txn}. We expect that it is due to the small $\Upsilon \mu^+ \mu^-$ branching ratio or lower $bb\bar b \bar b$
production rates relative to the charm sector. Finally, the $S-$wave spectra for $cc\bar c \bar c$ and $bb\bar b \bar b$
systems are plotted in Figure~\ref{mass} for reference.

\begin{table*}[!htb]
\begin{center}
\caption{\label{mass2} Predicted masses of radial excitations for $cc\bar c \bar c$ and $bb\bar b \bar b$ systems.}
\begin{tabular*}{18cm}{@{\extracolsep{\fill}}*{6}{p{2.7cm}<{\centering}}}
\hline\hline
System & $J^P$  & Configuration                                             & $\langle H\rangle$ (MeV) & Mass (MeV)  & Eigenvector\\\hline
 $cc\bar c \bar c$~(I) & $0^{++}$  &  $|\{cc\}^{\bar 3}_1 \{\bar c \bar c\}^3_1\rangle_0$
 & \multirow{2}{*}{$\begin{pmatrix}6917&-39 \\-39&6871\end{pmatrix}$}
              & \multirow{2}{*}{$\begin{bmatrix}6849 \\6940 \end{bmatrix}$}  & \multirow{2}{*}{$\begin{bmatrix}(0.500, 0.866)\\
               (0.866, -0.500)\end{bmatrix}$}\\
   &               &  $|\{cc\}^6_0 \{\bar c \bar c\}^{\bar 6}_0\rangle_0$    \\
 &$1^{+-}$  &  $|\{cc\}^{\bar 3}_1 \{\bar c \bar c\}^3_1\rangle_1$    & 6928  &  6928  &  1\\
 &$2^{++}$  &  $|\{cc\}^{\bar 3}_1 \{\bar c \bar c\}^3_1\rangle_2$    & 6948  &  6948  &  1\\\hline
 $cc\bar c \bar c$~(II) & $0^{++}$  &  $|\{cc\}^{\bar 3}_1 \{\bar c \bar c\}^3_1\rangle_0$
 & \multirow{2}{*}{$\begin{pmatrix}7046&-19 \\-19&7042\end{pmatrix}$}
              & \multirow{2}{*}{$\begin{bmatrix}7025 \\7063 \end{bmatrix}$}  & \multirow{2}{*}{$\begin{bmatrix}(0.664, 0.748)\\
               (0.748, -0.664)\end{bmatrix}$}\\
   &               &  $|\{cc\}^6_0 \{\bar c \bar c\}^{\bar 6}_0\rangle_0$    \\
 &$1^{+-}$  &  $|\{cc\}^{\bar 3}_1 \{\bar c \bar c\}^3_1\rangle_1$    & 7052  &  7052  &  1\\
 &$2^{++}$  &  $|\{cc\}^{\bar 3}_1 \{\bar c \bar c\}^3_1\rangle_2$    & 7064  &  7064  &  1\\\hline

 $bb\bar b \bar b$~(I) &$0^{++}$  &  $|\{bb\}^{\bar 3}_1 \{\bar b \bar b\}^3_1\rangle_0$     & \multirow{2}{*}{$\begin{pmatrix}19621
 & -14 \\-14&19571\end{pmatrix}$}
               & \multirow{2}{*}{$\begin{bmatrix}19567 \\19625 \end{bmatrix}$}  & \multirow{2}{*}{$\begin{bmatrix}(0.258, 0.966)\\
               (0.966, -0.258)\end{bmatrix}$}\\
    &             &  $|\{bb\}^6_0 \{\bar b \bar b\}^{\bar 6}_0\rangle_0$    \\
 &$1^{+-}$  &  $|\{bb\}^{\bar 3}_1 \{\bar b \bar b\}^3_1\rangle_1$    & 19625  &  19625  &  1\\
 &$2^{++}$  &  $|\{bb\}^{\bar 3}_1 \{\bar b \bar b\}^3_1\rangle_2$    & 19633  &  19633  &  1\\\hline
  $bb\bar b \bar b$~(II)  &$0^{++}$  &  $|\{bb\}^{\bar 3}_1 \{\bar b \bar b\}^3_1\rangle_0$     & \multirow{2}{*}{$\begin{pmatrix}19731
  & -6 \\-6&19736\end{pmatrix}$}
               & \multirow{2}{*}{$\begin{bmatrix}19726 \\19740 \end{bmatrix}$}  & \multirow{2}{*}{$\begin{bmatrix}(-0.822, -0.570)\\
               (0.570, -0.822)\end{bmatrix}$}\\
    &             &  $|\{bb\}^6_0 \{\bar b \bar b\}^{\bar 6}_0\rangle_0$    \\
 &$1^{+-}$  &  $|\{bb\}^{\bar 3}_1 \{\bar b \bar b\}^3_1\rangle_1$    & 19733  &  19733  &  1\\
 &$2^{++}$  &  $|\{bb\}^{\bar 3}_1 \{\bar b \bar b\}^3_1\rangle_2$    & 19736  &  19736  &  1\\
\hline\hline
\end{tabular*}
\end{center}
\end{table*}

\begin{figure}[htb]
\includegraphics[scale=1.0]{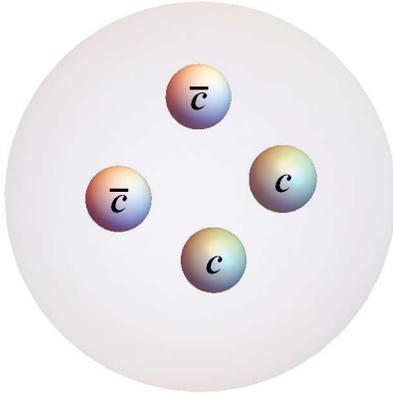}
\vspace{0.0cm} \caption{The narrow $cc \bar c \bar c$ structure near 6.9 GeV.}
\label{cccc}
\end{figure}

The proportions of the hidden color components and root mean square radii for excited $QQ\bar Q \bar Q$ states are listed in
Table~\ref{pro2}. The small root mean square radii for these states suggest that they are compact tetraquarks, and the sketch of the $cc \bar c \bar c$
structure near 6.9 GeV is presented in Figure~\ref{cccc}. For the excited
states, besides the $J/\psi J/\psi$ and $\Upsilon \Upsilon$ final states, lots of fall-apart channels are also open. The
possible decay modes via $S-$wave and $P-$wave are listed in Table~\ref{decay}.

\begin{table*}[!htb]
\begin{center}
\caption{\label{pro2} The color proportions and root mean square radii of the radial excited states for $cc\bar c \bar c$ and
$bb\bar b \bar b$ systems.  The units of masses and root mean square radii are in MeV and fm, respectively.}
\begin{tabular*}{18cm}{@{\extracolsep{\fill}}*{11}{p{1.3cm}<{\centering}}}
\hline\hline
 System  & $J^P$ & Mass  &   $|\bar 3 3\rangle$  &  $|6 \bar 6\rangle$  &  $|11\rangle$  & $|88\rangle$  &  $\langle \boldsymbol r_{12}^2
 \rangle^{1/2}$     &  $\langle \boldsymbol r^2 \rangle^{1/2}$  &
 $\langle \boldsymbol r_{13}^2  \rangle^{1/2}$
 &  $\langle \boldsymbol r^{\prime 2} \rangle^{1/2}$ \\\hline
 $cc\bar c \bar c$~(I)  & $0^{++}$ & 6849  & 25.0\%  &  75.0\% &  58.3\%  & 41.7\%   &  0.630  &  0.351     &  0.567  &  0.445  \\
  & $0^{++}$ & 6940  &   75.0\%  &  25.0\%  &  41.7\%  & 58.3\%   &  0.530  &  0.452     &  0.587  &  0.374  \\
  & $1^{+-}$ & 6928  &   100\%  &  0\%  &  33.3\%  & 66.7\%   &  0.471  &  0.504     &  0.604  &  0.333 \\
  & $2^{++}$ & 6948  &   100\%  &  0\%  &  33.3\%  & 66.7\%   &  0.468  &  0.520     &  0.616  &  0.331  \\\hline
  $cc\bar c \bar c$~(II)  & $0^{++}$ & 7025  & 44.0\%  &  56.0\% &  52.0\%  & 48.0\%   &  0.642  &  0.320     &  0.556  &  0.454  \\
  & $0^{++}$ & 7063  &   56.0\%  &  44.0\%  &  48.0\%  & 52.0\%   &  0.631  &  0.329     &  0.554  &  0.446 \\
  & $1^{+-}$ & 7052  &   100\%  &  0\%  &  33.3\%  & 66.7\%   &  0.592  &  0.361     &  0.553  &  0.419  \\
  & $2^{++}$ & 7064  &   100\%  &  0\%  &  33.3\%  & 66.7\%   &  0.598  &  0.365     &  0.558  &  0.423  \\\hline
 $bb\bar b \bar b$~(I)  & $0^{++}$ & 19567  &   6.7\%  & 93.3\%  &  64.4\%  & 35.6\%   &  0.436  &  0.208     &  0.372  &  0.309 \\
  & $0^{++}$ & 19625  &   93.3\%  &  6.7\%  &  35.6\%  & 64.4\%   &  0.324  &  0.334     &  0.405  &  0.229  \\
  & $1^{+-}$ & 19625  &   100\%  &  0\%  &  33.3\%  & 66.7\%   &  0.313  &  0.345     &  0.410  &  0.221   \\
  & $2^{++}$ & 19633  &   100\%  &  0\%  &  33.3\%  & 66.7\%   &  0.312  &  0.351     &  0.414  &  0.221 \\\hline
  $bb\bar b \bar b$~(II)  & $0^{++}$ & 19726  &   67.6\%  & 32.4\%  &  44.1\%  & 55.9\%   &  0.413  &  0.228     &  0.370  &  0.292 \\
  & $0^{++}$ & 19740  &   32.4\%  &  67.6\%  &  55.9\%  & 44.1\%   &  0.428  &  0.211     &  0.369  &  0.303  \\
  & $1^{+-}$ & 19733  &   100\%  &  0\%  &  33.3\%  & 66.7\%   &  0.398  &  0.244     &  0.372  &  0.282   \\
  & $2^{++}$ & 19736  &   100\%  &  0\%  &  33.3\%  & 66.7\%   &  0.399  &  0.245     &  0.374  &  0.282 \\

\hline\hline
\end{tabular*}
\end{center}
\end{table*}

\begin{table*}[htb]
\begin{center}
\fontsize{8.5pt}{24pt}
\caption{\label{decay} The decay channels of the $cc\bar c\bar c$ and $bb\bar b\bar b$ tetraquarks via fall-apart mechanism. }
\begin{tabular*}{18cm}{cccc}
\hline\hline
 System  & $J^P$ & $S-$wave  &  $P-$wave \\\hline
 $cc\bar c \bar c$  & $0^{++}$ & $\eta_c \eta_c$, $J/\psi J/\psi$, $\eta_c(2S) \eta_c$, $\psi(2S) J/\psi$, $h_c h_c$, $\chi_{c0}\chi_{c0}$,
 $\chi_{c1}\chi_{c1}$,  $\chi_{c2}\chi_{c2}$ & $\eta_c \chi_{c1}$, $J/\psi h_c$, $\eta_c(2S) \chi_{c1}$, $\psi(2S) h_c$ \\
  & $1^{+-}$ & $\eta_c J/\psi$, $\eta_c(2S) J/\psi$, $\eta_c \psi(2S)$, $h_c \chi_{c0}$,  $h_c \chi_{c1}$, $h_c \chi_{c2}$
  &   $\eta_c h_c$, $J/\psi \chi_{c0}$, $J/\psi \chi_{c1}$, $J/\psi \chi_{c2}$, $\eta_c(2S) h_c$, $\psi(2S) \chi_{c0}$,
  $\psi(2S) \chi_{c1}$, $\psi(2S) \chi_{c2}$  \\
  & $2^{++}$ & $J/\psi J/\psi$, $\psi(2S) J/\psi$, $h_c h_c$, $\chi_{c1}\chi_{c1}$,  $\chi_{c2}\chi_{c2}$  &  $\eta_c \chi_{c1}$,
  $\eta_c \chi_{c2}$, $J/\psi h_c$, $\eta_c(2S) \chi_{c1}$, $\eta_c(2S) \chi_{c2}$, $\psi(2S) h_c$   \\\hline
 $bb\bar b \bar b$  & $0^{++}$ & $\eta_b \eta_b$, $\Upsilon \Upsilon$, $\eta_b(2S) \eta_b$, $\Upsilon(2S) \Upsilon$, $h_b h_b$,
 $\chi_{b0}\chi_{b0}$, $\chi_{b1}\chi_{b1}$,  $\chi_{b2}\chi_{b2}$ & $\eta_b \chi_{b1}$, $\Upsilon h_b$, $\eta_b(2S) \chi_{b1}$,
 $\Upsilon(2S) h_b$ \\
  & $1^{+-}$ & $\eta_b \Upsilon$, $\eta_b(2S) \Upsilon$, $\eta_b \Upsilon(2S)$, $h_b \chi_{b0}$,  $h_b \chi_{b1}$, $h_b \chi_{b2}$
  &   $\eta_b h_b$, $\Upsilon \chi_{b0}$, $\Upsilon \chi_{b1}$, $\Upsilon \chi_{b2}$, $\eta_b(2S) h_b$, $\Upsilon(2S) \chi_{b0}$,
  $\Upsilon(2S) \chi_{b1}$, $\Upsilon(2S) \chi_{b2}$ \\
  & $2^{++}$ & $\Upsilon \Upsilon$, $\Upsilon(2S) \Upsilon$, $h_b h_b$, $\chi_{b1}\chi_{b1}$,  $\chi_{b2}\chi_{b2}$
  &  $\eta_b \chi_{b1}$, $\eta_b \chi_{b2}$, $\Upsilon h_b$, $\eta_b(2S) \chi_{b1}$, $\eta_b(2S) \chi_{b2}$, $\Upsilon(2S) h_b$   \\
\hline\hline
\end{tabular*}
\end{center}
\end{table*}

Since the structure near 6.9 GeV corresponds to a radial excited state in our calculation, the $\psi(2S) J/\psi$ channel is expected
to be significant. The ratios
\begin{equation}
R=\frac{\Gamma[cc\bar c \bar c \to J/\psi J/\psi ]}{\Gamma[cc\bar c \bar c \to \psi(2S) J/\psi ]}
\end{equation}
can be adopted to describe the relative magnitudes between $J/\psi J/\psi$ and $\psi(2S) J/\psi$ final states. For simplicity, one can
assume the decay amplitudes are proportional to the overlap of initial and final states, and the proportional coefficient can be canceled in the final ratios. Here, the wave functions for initial tetraquarks have been obtained by solving the generalized eigenvalue problem, and the wave functions of $J/\psi$, $\psi(2S)$, $\Upsilon$, and $\Upsilon(2S)$ can be got within the
relativized quark model as well. With these wave functions, the ratios for $0^{++}$ and $2^{++}$ states can be estimated to be
\begin{equation}
R[cc\bar c \bar c(6849)]=0.113,
\end{equation}
\begin{equation}
R[cc\bar c \bar c(6940)]=0.122,
\end{equation}
\begin{equation}
R[cc\bar c \bar c(6948)]=0.075.
\end{equation}
Combined with the branching ratios of $J/\psi \to \mu^+ \mu^-$ and $\psi(2S) \to \mu^+ \mu^-$, one can further define
\begin{equation}
R_{4\mu} = \frac{\Gamma[cc\bar c \bar c \to J/\psi J/\psi \to \mu^+ \mu^- \mu^+ \mu^-]}{\Gamma[cc\bar c \bar c \to \psi(2S) J/\psi
\to \mu^+ \mu^- \mu^+ \mu^- ]}.
\end{equation}
Then, the ratios $R_{4\mu}$ are predicted to be
\begin{equation}
R_{4\mu}[cc\bar c \bar c(6849)]=0.843,
\end{equation}
\begin{equation}
R_{4\mu}[cc\bar c \bar c(6940)]=0.910,
\end{equation}
\begin{equation}
R_{4\mu}[cc\bar c \bar c(6948)]=0.559.
\end{equation}
It can be found that the $\psi(2S) J/\psi$ channel for the excited states is important even though the phase spaces are smaller.
The similar situation occurs for the lower excited $bb\bar b \bar b$ states, where the $R_{4\mu}$ of $bb\bar b \bar b(19567)$,
$bb\bar b \bar b(19625)$, and $bb\bar b \bar b(19633)$ states are 0.113, 0.111, and 0.084, respectively. These ratios indicate that the
lower excited $bb\bar b \bar b$ states can decay to $\mu^+ \mu^- \mu^+ \mu^-$ final states through $\Upsilon(2S) \Upsilon$ more easily than $\Upsilon \Upsilon$ mode. Future experiments can search
for these states in $\psi(2S) J/\psi$ and $\Upsilon(2S) \Upsilon$ final states.

\section{Summary}{\label{Summary}}

In this work, we investigate the masses of fully heavy tetraquarks $cc \bar c \bar c$ and $bb \bar b \bar b$ in an extended relativized quark model.
The four-body Hamiltonian including the Coulomb potential, confining potential, spin-spin interactions, and relativistic corrections are
solved within the variational method. Our estimations indicate that the broad structure around 6.4 GeV should contain one or more ground $cc \bar c \bar c$ tetraquark states, while the narrow structure near 6.9 GeV can be categorized as the first radial excitation of $cc \bar c \bar c$ system. The significant hidden color component and
small root mean square radii demonstrate that these states are compact tetraquarks. For the radial excited states, the decay ratios
between the $J/\psi J/\psi$ and $\psi(2S) J/\psi$ [or $\Upsilon \Upsilon$ and $\Upsilon(2S) \Upsilon$] modes are also qualitatively discussed with the wave functions of the tetraquarks and mesons. Our results
show that the $\psi(2S) J/\psi$ or $\Upsilon(2S) \Upsilon$ channel is significant for these excited tetraquarks. We hope
our sophisticated calculations of the fully heavy tetraquarks may provide valuable information for future experimental searches.

\bigskip
\noindent
\begin{center}
{\bf ACKNOWLEDGEMENTS}\\
\end{center}
This project is supported by the National Natural Science Foundation of China under Grants No.~11705056, No.~11775050, No.~11947224,
No.~11975245, and No.~U1832173, by the fund provided to the Sino-German CRC 110 ``Symmetries and the Emergence of Structure in QCD"
project by the NSFC under Grant No.~11621131001,
and by the Key Research Program of Frontier Sciences, CAS, Grant No. Y7292610K1.


\begin{thebibliography}{110}

\bibitem{Choi:2003ue}
S.~Choi \textit{et al.} [Belle],
Observation of a narrow charmonium - like state in exclusive $B^{\pm} \to  K^{\pm} \pi^+ \pi^- J / \psi$ decays,
Phys. Rev. Lett. \textbf{91}, 262001 (2003).
%doi:10.1103/PhysRevLett.91.262001
%[arXiv:hep-ex/0309032 [hep-ex]].
%1787 citations counted in INSPIRE as of 25 Jun 2020

\bibitem{Klempt:2007cp} 
  E.~Klempt and A.~Zaitsev,
  Glueballs, Hybrids, Multiquarks. Experimental facts versus QCD inspired concepts,
  Phys.\ Rept.\  {\bf 454}, 1 (2007).
  %doi:10.1016/j.physrep.2007.07.006
  %[arXiv:0708.4016 [hep-ph]].
  %%CITATION = doi:10.1016/j.physrep.2007.07.006;%%
  %669 citations counted in INSPIRE as of 09 Feb 2020

\bibitem{Brambilla:2010cs}
  N.~Brambilla {\it et al.},
  Heavy Quarkonium: Progress, Puzzles, and Opportunities,
  Eur.\ Phys.\ J.\ C {\bf 71}, 1534 (2011).
  %%doi:10.1140/epjc/s10052-010-1534-9
  %%[arXiv:1010.5827 [hep-ph]].
  %%CITATION = doi:10.1140/epjc/s10052-010-1534-9;%%
  %1444 citations counted in INSPIRE as of 07 Feb 2020

\bibitem{Chen:2016qju}
  H.~X.~Chen, W.~Chen, X.~Liu and S.~L.~Zhu,
  The hidden-charm pentaquark and tetraquark states,
  Phys.\ Rept.\  {\bf 639},1 (2016). 
  %%doi:10.1016/j.physrep.2016.05.004
  %%[arXiv:1601.02092 [hep-ph]].
  %%CITATION = doi:10.1016/j.physrep.2016.05.004;%%
  %499 citations counted in INSPIRE as of 07 Feb 2020	
  
\bibitem{Lebed:2016hpi}
  R.~F.~Lebed, R.~E.~Mitchell and E.~S.~Swanson,
  Heavy-Quark QCD Exotica,
  Prog.\ Part.\ Nucl.\ Phys.\  {\bf 93}, 143 (2017).
  %%doi:10.1016/j.ppnp.2016.11.003
  %%[arXiv:1610.04528 [hep-ph]].
  %%CITATION = doi:10.1016/j.ppnp.2016.11.003;%%
  %226 citations counted in INSPIRE as of 07 Feb 2020  

\bibitem{Guo:2017jvc}
  F.~K.~Guo, C.~Hanhart, U.~G.~Mei\ss ner, Q.~Wang, Q.~Zhao and B.~S.~Zou,
  Hadronic molecules,
  Rev.\ Mod.\ Phys.\  {\bf 90}, 015004 (2018). 
  %%doi:10.1103/RevModPhys.90.015004
  %%[arXiv:1705.00141 [hep-ph]].
  %%CITATION = doi:10.1103/RevModPhys.90.015004;%%
  %349 citations counted in INSPIRE as of 07 Feb 2020
  
\bibitem{Esposito:2016noz}
  A.~Esposito, A.~Pilloni and A.~D.~Polosa,
  Multiquark Resonances,
  Phys.\ Rept.\  {\bf 668}, 1 (2017).
  %%doi:10.1016/j.physrep.2016.11.002
  %%[arXiv:1611.07920 [hep-ph]].
  %%CITATION = doi:10.1016/j.physrep.2016.11.002;%%
  %257 citations counted in INSPIRE as of 07 Feb 2020
  
\bibitem{Ali:2017jda}
  A.~Ali, J.~S.~Lange and S.~Stone,
  Exotics: Heavy Pentaquarks and Tetraquarks,
  Prog.\ Part.\ Nucl.\ Phys.\  {\bf 97}, 123 (2017).
  %%doi:10.1016/j.ppnp.2017.08.003
  %%[arXiv:1706.00610 [hep-ph]].
  %%CITATION = doi:10.1016/j.ppnp.2017.08.003;%%
  %190 citations counted in INSPIRE as of 07 Feb 2020



\bibitem{Liu:2019zoy}
  Y.~R.~Liu, H.~X.~Chen, W.~Chen, X.~Liu and S.~L.~Zhu,
  Pentaquark and Tetraquark states,
  Prog.\ Part.\ Nucl.\ Phys.\  {\bf 107}, 237 (2019).
  %%doi:10.1016/j.ppnp.2019.04.003
  %%[arXiv:1903.11976 [hep-ph]].
  %%CITATION = doi:10.1016/j.ppnp.2019.04.003;%%
  %77 citations counted in INSPIRE as of 07 Feb 2020	


\bibitem{Brambilla:2019esw}
  N.~Brambilla, S.~Eidelman, C.~Hanhart, A.~Nefediev, C.~P.~Shen, C.~E.~Thomas, A.~Vairo and C.~Z.~Yuan,
  The $XYZ$ states: experimental and theoretical status and perspectives,
  arXiv:1907.07583.
  %%CITATION = ARXIV:1907.07583;%%
  %50 citations counted in INSPIRE as of 07 Feb 2020
  
\bibitem{Dong:2017gaw}
  Y.~Dong, A.~Faessler and V.~E.~Lyubovitskij,
  Description of heavy exotic resonances as molecular states using phenomenological Lagrangians,
  Prog.\ Part.\ Nucl.\ Phys.\  {\bf 94}, 282 (2017).
  %%doi:10.1016/j.ppnp.2017.01.002
  %%CITATION = doi:10.1016/j.ppnp.2017.01.002;%%
  %22 citations counted in INSPIRE as of 07 Feb 2020

\bibitem{Khachatryan:2016ydm}
V.~Khachatryan \textit{et al.} (CMS Collaboration),
Observation of $\Upsilon$(1S) pair production in proton-proton collisions at $ \sqrt{s}=8 $ TeV,
J. High Energy Phys. \textbf{05}, 013 (2017).
%%doi:10.1007/JHEP05(2017)013
%%[arXiv:1610.07095 [hep-ex]].
%59 citations counted in INSPIRE as of 25 Jun 2020

\bibitem{CMSMeeting}
CMS Collaboration Collaboration, S. Durgut for the collaboration.
https://meetings.aps.org/Meeting/APR18/Session/U09.6


\bibitem{Yi:2018fxo}
K.~Yi,
Things that go bump in the night: From $J/\psi\phi$ to other mass spectrum,
Int. J. Mod. Phys. A \textbf{33}, 1850224 (2019). 
%%doi:10.1142/S0217751X1850224X
%%[arXiv:1806.08398 [hep-ph]].
%4 citations counted in INSPIRE as of 25 Jun 2020


\bibitem{Aaij:2018zrb}
  R.~Aaij {\it et al.} (LHCb Collaboration),
  Search for beautiful tetraquarks in the $\Upsilon(1S)\mu\mu$ invariant-mass spectrum,
  J. High Energy Phys.{\bf 10}, 086 (2018).
  %%CITATION = ARXIV:1806.09707;%%

\bibitem{Sirunyan:2020txn}
A.~M.~Sirunyan \textit{et al.} (CMS Collaboration),
Measurement of the $\Upsilon$(1S) pair production cross section and search for resonances decaying to $\Upsilon(1S) \mu^+\mu^-$ in proton-proton collisions at $\sqrt{s}=$ 13 TeV,
arXiv:2002.06393.
%2 citations counted in INSPIRE as of 20 May 2020

\bibitem{Wang:2017jtz}
  Z.~G.~Wang,
  Analysis of the $QQ\bar{Q}\bar{Q}$ tetraquark states with QCD sum rules,
  Eur.\ Phys.\ J.\ C {\bf 77}, 432 (2017).
  %doi:10.1140/epjc/s10052-017-4997-0
  %[arXiv:1701.04285 [hep-ph]].
  %%CITATION = doi:10.1140/epjc/s10052-017-4997-0;%%
  %20 citations counted in INSPIRE as of 06 Jul 2018


\bibitem{Karliner:2016zzc}
  M.~Karliner, S.~Nussinov, and J.~L.~Rosner,
  $Q Q \bar Q \bar Q$ states: Masses, production, and decays,
  Phys.\ Rev.\ D {\bf 95}, 034011 (2017).
  %doi:10.1103/PhysRevD.95.034011
 % [arXiv:1611.00348 [hep-ph]].
  %%CITATION = doi:10.1103/PhysRevD.95.034011;%%
  %32 citations counted in INSPIRE as of 06 Jul 2018

\bibitem{Berezhnoy:2011xn}
  A.~V.~Berezhnoy, A.~V.~Luchinsky and A.~A.~Novoselov,
  Tetraquarks composed of 4 heavy quarks,
  Phys.\ Rev.\ D {\bf 86}, 034004 (2012).
  %doi:10.1103/PhysRevD.86.034004
%  [arXiv:1111.1867 [hep-ph]].
  %%CITATION = doi:10.1103/PhysRevD.86.034004;%%
  %24 citations counted in INSPIRE as of 06 Jul 2018

%\cite{Bai:2016int}
\bibitem{Bai:2016int}
Y.~Bai, S.~Lu and J.~Osborne,
Beauty-full Tetraquarks,
Phys. Lett. B \textbf{798}, 134930 (2019).
%doi:10.1016/j.physletb.2019.134930
%[arXiv:1612.00012 [hep-ph]].
%40 citations counted in INSPIRE as of 25 Jun 2020

\bibitem{Anwar:2017toa}
  M.~N.~Anwar, J.~Ferretti, F.~K.~Guo, E.~Santopinto, and B.~S.~Zou,
  Spectroscopy and decays of the fully-heavy tetraquarks,
  Eur.\ Phys.\ J.\ C {\bf 78}, 647 (2018).
%  doi:10.1140/epjc/s10052-018-6073-9
%  [arXiv:1710.02540 [hep-ph]].
  %%CITATION = doi:10.1140/epjc/s10052-018-6073-9;%%
  %16 citations counted in INSPIRE as of 03 Jan 2019


\bibitem{Esposito:2018cwh}
  A.~Esposito and A.~D.~Polosa,
  A $bb\bar b\bar b$ di-bottomonium at the LHC,
  Eur Phys J.C {\bf 78}, 782 (2018).
  %%CITATION = ARXIV:1807.06040;%%

\bibitem{Chen:2016jxd}
  W.~Chen, H.~X.~Chen, X.~Liu, T.~G.~Steele and S.~L.~Zhu,
  Hunting for exotic doubly hidden-charm/bottom tetraquark states,
  Phys.\ Lett.\ B {\bf 773}, 247 (2017).
  %doi:10.1016/j.physletb.2017.08.034
 % [arXiv:1605.01647 [hep-ph]].
  %%CITATION = doi:10.1016/j.physletb.2017.08.034;%%
  %20 citations counted in INSPIRE as of 06 Jul 2018

%%%%%%%%%%%%%%%%%%%%%%%%%%%%%%%%%%%%%%%%%%%%%%%%%%

\bibitem{Debastiani:2017msn}
  V.~R.~Debastiani and F.~S.~Navarra,
  A non-relativistic model for the $[cc][\bar{c}\bar{c}]$ tetraquark,
  Chin.Phys.C {\bf 43}, 013105 (2018).
  %%CITATION = ARXIV:1706.07553;%%
  %5 citations counted in INSPIRE as of 06 Jul 2018

%\cite{Wang:2018poa}
\bibitem{Wang:2018poa}
Z.~G.~Wang and Z.~Y.~Di,
Analysis of the vector and axialvector $QQ\bar{Q}\bar{Q}$ tetraquark states with QCD sum rules,
Acta Phys. Polon. B \textbf{50}, 1335 (2019).
%doi:10.5506/APhysPolB.50.1335
%[arXiv:1807.08520 [hep-ph]].
%9 citations counted in INSPIRE as of 25 Jun 2020


\bibitem{Wu:2016vtq}
  J.~Wu, Y.~R.~Liu, K.~Chen, X.~Liu, and S.~L.~Zhu,
  Heavy-flavored tetraquark states with the $QQ\bar{Q}\bar{Q}$ configuration,
  Phys.\ Rev.\ D {\bf 97}, 094015 (2018).
  %doi:10.1103/PhysRevD.97.094015
 % [arXiv:1605.01134 [hep-ph]].
  %%CITATION = doi:10.1103/PhysRevD.97.094015;%%
  %22 citations counted in INSPIRE as of 06 Jul 2018


\bibitem{Lloyd:2003yc}
  R.~J.~Lloyd and J.~P.~Vary,
  All charm tetraquarks,
  Phys.\ Rev.\ D {\bf 70}, 014009 (2004).
%  doi:10.1103/PhysRevD.70.014009
%  [hep-ph/0311179].
  %%CITATION = doi:10.1103/PhysRevD.70.014009;%%
  %28 citations counted in INSPIRE as of 06 Jul 2018

\bibitem{Ader:1981db}
  J.~P.~Ader, J.~M.~Richard and P.~Taxil,
  Do narrow heavy multi - quark states exist,
  Phys.\ Rev.\ D {\bf 25}, 2370 (1982).
  %doi:10.1103/PhysRevD.25.2370
  %%CITATION = doi:10.1103/PhysRevD.25.2370;%%
  %140 citations counted in INSPIRE as of 06 Jul 2018

\bibitem{Hughes:2017xie}
  C.~Hughes, E.~Eichten, and C.~T.~H.~Davies,
  Searching for beauty-fully bound tetraquarks using lattice nonrelativistic QCD,
  Phys.\ Rev.\ D {\bf 97}, 054505 (2018).
  %doi:10.1103/PhysRevD.97.054505
%  [arXiv:1710.03236 [hep-lat]].
  %%CITATION = doi:10.1103/PhysRevD.97.054505;%%
  %9 citations counted in INSPIRE as of 06 Jul 2018



%%%%%%%%%%%%%%%%%%%%%%%%%%%%%%%%%%%

\bibitem{Richard:2018yrm}
  J.~M.~Richard, A.~Valcarce, and J.~Vijande,
  Few-body quark dynamics for doubly heavy baryons and tetraquarks,
  Phys.\ Rev.\ C {\bf 97}, 035211 (2018).
%  doi:10.1103/PhysRevC.97.035211
%  [arXiv:1803.06155 [hep-ph]].
  %%CITATION = doi:10.1103/PhysRevC.97.035211;%%
  %1 citations counted in INSPIRE as of 24 Jul 2018

\bibitem{Liu:2019zuc}
M.~S.~Liu, Q.~F.~L\"u, X.~H.~Zhong and Q.~Zhao,
All-heavy tetraquarks,
Phys. Rev. D \textbf{100}, 016006 (2019). 
%%doi:10.1103/PhysRevD.100.016006
%%[arXiv:1901.02564 [hep-ph]].
%20 citations counted in INSPIRE as of 25 Jun 2020

\bibitem{Wang:2019rdo}
G.~J.~Wang, L.~Meng and S.~L.~Zhu,
Spectrum of the fully-heavy tetraquark state $QQ\bar Q' \bar Q'$,
Phys. Rev. D \textbf{100}, 096013 (2019).
%%doi:10.1103/PhysRevD.100.096013
%%[arXiv:1907.05177 [hep-ph]].
%11 citations counted in INSPIRE as of 25 Jun 2020

%\cite{Chen:2019dvd}
\bibitem{Chen:2019dvd}
X.~Chen,
Analysis of hidden-bottom $bb\bar{b}\bar{b}$ states,
Eur. Phys. J. A \textbf{55}, 106 (2019).
%doi:10.1140/epja/i2019-12807-2
%[arXiv:1902.00008 [hep-ph]].
%8 citations counted in INSPIRE as of 25 Jun 2020

\bibitem{Deng:2020iqw}
C.~R.~Deng, H.~Chen and J.~L.~Ping,
Towards the understanding of fully-heavy tetraquark states from various models,
arXiv:2003.05154.
%0 citations counted in INSPIRE as of 20 May 2020

\bibitem{LHCb:2020}
L. An, LHC Seminar,
https://indico.cern.ch/event/900972/.

\bibitem{liu:2020eha}
M.~S.~liu, F.~X.~Liu, X.~H.~Zhong and Q.~Zhao,
Full-heavy tetraquark states and their evidences in the LHCb di-$J/\psi$ spectrum,
arXiv:2006.11952.
%2 citations counted in INSPIRE as of 25 Jun 2020

\bibitem{Wang:2020ols}
Z.~G.~Wang,
Tetraquark candidates in the LHCb's di-$J/\psi$ mass spectrum,
arXiv:2006.13028.
%0 citations counted in INSPIRE as of 25 Jun 2020

%\cite{1802716}
\bibitem{1802716}
X.~Jin, Y.~Xue, H.~Huang and J.~Ping,
Full-heavy tetraquarks in constituent quark models,
arXiv:2006.13745.
%0 citations counted in INSPIRE as of 25 Jun 2020

\bibitem{1802721}
G.~Yang, J.~Ping, L.~He and Q.~Wang,
A potential model prediction of fully-heavy tetraquarks $QQ\bar{Q}\bar{Q}$ ($Q=c, b$),
arXiv:2006.13756.
%0 citations counted in INSPIRE as of 25 Jun 2020

\bibitem{Lu:2020rog}
Q.~F.~L\"u, D.~Y.~Chen and Y.~B.~Dong,
Masses of doubly heavy tetraquarks $T_{QQ^\prime}$ in a relativized quark model,
arXiv:2006.08087.
%0 citations counted in INSPIRE as of 17 Jun 2020

\bibitem{Godfrey:1985xj}
  S.~Godfrey and N.~Isgur,
  Mesons in a relativized quark model with chromodynamics,
  Phys.\ Rev.\ D {\bf 32}, 189 (1985).
  %%CITATION = PHRVA,D32,189;%%
  %2081 citations counted in INSPIRE as of 13 Oct 2015

\bibitem{Hiyama:2003cu}
  E.~Hiyama, Y.~Kino, and M.~Kamimura,
  Gaussian expansion method for few-body systems,
  Prog.\ Part.\ Nucl.\ Phys.\  {\bf 51}, 223 (2003).
 % doi:10.1016/S0146-6410(03)90015-9
  %%CITATION = doi:10.1016/S0146-6410(03)90015-9;%%
  %333 citations counted in INSPIRE as of 10 Apr 2019


\bibitem{Richard:2019cmi}
J.~M.~Richard, A.~Valcarce and J.~Vijande,
Hall-Post inequalities: Review and application to molecules and tetraquarks,
Annals Phys. \textbf{412}, 168009 (2020).
%doi:10.1016/j.aop.2019.168009
%[arXiv:1910.08295 [nucl-th]].
%0 citations counted in INSPIRE as of 20 May 2020







\end{thebibliography}
\end{document}